\documentclass[aps,pre,reprint,amsmath,amssymb, superscriptaddress]{revtex4-1}
\usepackage[utf8]{inputenc}
\usepackage{amsmath}
\usepackage{amssymb}
\usepackage{hyperref}
\usepackage{color}
\usepackage{dcolumn}
\usepackage{bm}
\usepackage{graphicx}

\newcommand{\beq}{\begin{equation}}
\newcommand{\eeq}{\end{equation}}

\begin{document}

\title{Systemic risk measured by systems resiliency to initial shocks}

\author{Luka Klin\v{c}i\'c}
\affiliation{Department of Physics, Faculty of Science, University of Zagreb, Bijeni\v{c}ka c. 32, 10000 Zagreb, Croatia}
\author{Vinko Zlati\'c}
\affiliation{Theoretical Physics Division, Rudjer Bo\v{s}kovi\'c Institute, Bijeni\v{c}ka c. 54, 10000 Zagreb, Croatia}
\author{Guido Caldarelli}
\affiliation{DSMN University of Venice Ca'Foscari, Via Torino 155, 30171 Venezia Mestre, Italy}
\affiliation{ISC-CNR, Dipartimento di Fisica Univ. Sapienza, Rome, Italy}
\affiliation{London Institute for Mathematical Sciences, London, UK}
\affiliation{Fondazione per il futuro delle citt\`a  Florence, Italy}
\author{Hrvoje \v{S}tefan\v{c}i\'c}
\affiliation{Catholic University of Croatia, Ilica 242, 10000, Zagreb, Croatia}

\date{\today}

\begin{abstract}
    The study of systemic risk is often presented through the analysis of several measures referring to quantities used by practitioners and policy makers. Almost invariably, those measures  evaluate the size of the impact that exogenous events can exhibit on a financial system without analysing the nature of initial shock. Here we present a symmetric approach and propose a set of measures that are based on the amount of exogenous shock that can be absorbed by the system before it starts to deteriorate. For this purpose, we use a linearized version of DebtRank that allows to clearly show the onset of financial distress towards a correct systemic risk estimation. We show how we can explicitly compute localized and uniform exogenous shocks and explained their behavior though spectral graph theory. We also extend analysis to heterogeneous shocks that have to be computed by means of Monte Carlo simulations. We believe that our approach is more general and natural and allows to express in a standard way the failure risk in financial systems.
\end{abstract}

\maketitle

\section{Introduction}
The study of networks in nature and society has led to a number of interesting models both aimed at explaining the topology and both explaining  dynamics running on them\cite{NewmanNetworks}. 
Within these studies of dynamical models, the ones related to the systemic breakdown of the system represent a broad and an important direction. A special focus within this class of models is given to the model of systemic risk in economical i.e. financial systems \cite{chinazzi2015systemic, gai2019networks, bardoscia2021physics}.

Systemic risk, or the risk of collapse for the major part of a complex system, has been a major concern for academics and an important issue for policymakers and regulators. One approach to analyzing systemic risk is to use complexity theory and network analysis \cite{battiston2016complexity}. This allows to study the interconnections and dependencies between financial institutions. Studies have shown that the network structure of the financial system can have a significant impact on the transmission of shocks and the likelihood of contagion. A recent study \cite{gai2010contagion} has shown that a network model can be used to measure and manage systemic risk in the interbank market (hereafter ``interbank network''). Similarly, it has been reported \cite{beale2011individual} that a densely interconnected financial system is associated with a higher level of systemic risk. 
This intuition resulted in more quantitative analysis on specific cases of interest
\cite{battiston2012debtrank,battiston2012liaisons,minoiu2013network,chinazzi2013post}. Recent research of systemic risk extends the financial network to multi-layer networks, including different assets and types of loans etc.~\cite{poledna2015multi,montagna2016multi, brummitt2015cascades}. 
These studies have provided some of the first evidence about the importance of network analysis in understanding and mitigating systemic risk and have been applied to inform regulators \cite{battiston2012debtrank}, in order to individually asses a systemic risk \cite{poledna2018identifying}, to simulate different policies, such as bank taxation \cite{zlatic2015reduction,poledna2016elimination}, or to understand which network architecures are more and which are less risky~\cite{krause2021controlling}.

Another important aspect of analyzing systemic risk is the use of debt ranking~\cite{battiston2012debtrank}, which is a method of assigning a relative importance or "rank" to different financial institutions based on their level of debt and their position in the network. Debt ranking, as other centrality measures, can be used to identify the most systemically important institutions, which are often referred to as "too-big-to-fail" or "to-connected-to-fail" institutions. These institutions are considered to be the most critical to the stability of the financial system and are therefore subject to more strict regulations and oversight. 

Aforementioned measures of networked systemic risk are all based on the consequence for financial system after a systemic event and propagation of default cascade through the financial network \cite{roukny2013default}. 

In this paper we propose a different approach to measure systemic risk as an \emph{amount of external shock the system can absorb before a systemic event takes place}. As a systemic event, we consider any default of financial institution that was caused by the shock propagation through the network. In technical terms the measure of systemic risk can be computed as an inverse problem of estimation of the initial conditions for the cases in which systemic event in the models took place. Even if this approach might look more complicated than the one usually studied, nevertheless through a model of networked systemic risk, {\em i.e.} the DebtRank\cite{bardoscia2015}, we can get some insight in the phenomenon and become able to provide quantitative measures that generalize the risk analysis to all the possible cases of exogenous shock.

In section 2 we present the version of DebtRank that we use in our analysis, and explain why it is a good model for dealing with the inverse problem. In the section 3 we describe the 3 different versions of shocks that can be evaluated using this method - namely uniform, localised and heterogenous shocks, from which we develop three new measures of systemic risk. In section 4 we present the results of the analysis. First  we present analytical results on very small networks; later, the simulated results for larger networks.
In the end we give conclusion and outline further research directions.

\section{DebtRank measure of systemic risk}
As a starting point in this analysis we will assume a complete knowledge of bank balances and interbank loans. The bank balance consists of assets - $A$, liabilities - $L$ and equity - $E$, and their relationship is a common balance sheet equation:
\beq
A=L+E\label{BalanceEquation}
\eeq

The systemic risk will be modelled using the DebtRank algorithm \cite{battiston2012debtrank}, more precisely its version published in \cite{bardoscia2015}. In this version, the system consists of $n$ banks that are represent by $n$ vertices in the network representing financial system. The most common mode of interaction between the banks are interbank loans, that are in the network represented by weighted edges $A_{ij}$, in which weights represent the sum of loans that the bank $i$ used to lend to bank $j$. Total network assets of bank $i$ that represents the amount of lending to other banks is $A_i=\sum_{j} A_{ij}$. The rest of the bank $i$ assets we denote as $A^E_i$. Every amount of lended assets $A_{ij}$ of bank $i$ is matched by the amount of liabilities $L_{ij}$ of the bank $j$ that needs to repay it at some time in the future. Similar to assets, liabilities of the bank $i$ which are not related to interbank finacial system are denoted as  s $L^E_i$. The equity of the bank $i$ is computed using the \ref{BalanceEquation} as $E_i = A_i - L_i$. In this paper we use convention that the bank $i$ becomes bankrupt in a case in which liabilities become larger than assets i.e. in the case when  $E_i \leq 0$.

To study the evolution of the system, we consider the set on non-bankrupted banks at time $t$
\begin{equation}
\mathcal{A}(t) = \{j : E_j(t) > 0\}\, .
\end{equation}

Further assumption is that when the bank $j$ goes bankrupt, its assets are taken from the system i.e. $A_{ij} = 0$, but liabilities remain $L_{ij}$ constant. This leads to temporal balance equation
\begin{equation}
E_i(t) = A^E_i(t) - L^E_i(t) + \sum_{j \in \mathcal{A}(t-1)} A_{ij}(t) -\sum^n_{j = 1}L_{ij}(t) \, ,\label{balanceEq2}
\end{equation}
in which the sum of assets includes only the banks that were ''healthy'' at time $t-1$, i.e. we assume that the information of bankruptcy takes one unit of time.
 
 This leads to the following mechanism of the propagation of financial shock, under the assumption that the assets of lender are changed proportionally to the equity of borrower in the previous time stamp i.e.  
 \begin{equation}
 A_{ij}(t+1) = 
 \begin{cases}
 A_{ij}(t)\frac{E_j(t)}{E_j(t-1)}\, , & j \in \mathcal{A}(t-1) \\
  A_{ij}(t) = 0\, , & j \not\in \mathcal{A}(t-1) \, .
 \end{cases}\label{cases}
 \end{equation}
Time evolution of assets can be computed using equations~\ref{balanceEq2} and ~\ref{cases} to get

\begin{equation}
E_i(t+1) - E_i(t) = \sum_{j \in \mathcal{A}(t-1)} \frac{A_{ij}(0)}{E_j(0)}\big( E_j(t) - E_j(t-1)  \big) \, ,
\end{equation}
where the equation \ref{balanceEq2} was used recursively., and $A_{ij}(1) = A_{ij}(0)$.

Evolution of equity can be written as
\begin{equation}
E_i(t+1) = \max  \Bigg[0,\, E_i(t) + \sum_{j = 1}^n \tilde{\Lambda }_{ij}(t) \big( E_j(t) - E_j(t-1)  \big) \Bigg]\, ,
\end{equation}
in which
\begin{equation}
 \tilde{\Lambda }_{ij}(t) =
 \begin{cases}
 \frac{A_{ij}(0)}{E_j(0)}\, , & j \in \mathcal{A}(t-1) \\
   0\, , & j \not\in \mathcal{A}(t-1)
 \end{cases} \, .\label{EquityEvolution}
 \end{equation}

Here maximum ensures that the equity after the bankruptcy can not become negative. Now we define financial shock on bank $i$ as 
\begin{equation}
h_i(t) = \frac{E_i(0) - E_i(t)}{E_i(0)}
\end{equation}
and using also \ref{EquityEvolution} the evolution of financial shock is
\begin{equation}
h_i(t+1) = \min  \Bigg[0,\, h_i(t) + \sum_{j = 1}^n \Lambda_{ij}(t) \big( h_j(t) - h_j(t-1)  \big) \Bigg]\, .\label{FinShockEvolution}
\end{equation}
in which
\begin{equation}
 \Lambda_{ij}(t) =
 \begin{cases}
 \frac{A_{ij}(0)}{E_i(0)}\, , & j \in \mathcal{A}(t-1) \\
   0\, , & j \not\in \mathcal{A}(t-1)
 \end{cases} \, .
 \end{equation}

Key component of DebtRank dynamics is a fact the system stability is completely determined by the properties of matrix $\Lambda(t)$.

If we consider the time period between two bankruptcies i.e. a period in which the matrix $\Lambda(t) = \Lambda$ is constant. Defining the change of financial shock as $\Delta h(t) = h(t) - h(t-1)$ and using the fact that $h(0) = 0$, the equation \ref{FinShockEvolution} can be written in a matrix form

\begin{equation}
\begin{split}
\Delta h(t+1) &= \Lambda \Delta h(t) \\ 
&= \Lambda^t \Delta h(1) = \Lambda^t h(1)\, ,
\end{split}
\end{equation}
 The financial shock at $t + 1$, is a sum of previous shocks
\begin{equation}
h(t+1) = \sum^{t + 1}_{t' = 0} \Delta h(t') = \sum^{t + 1}_{t' = 0} \Lambda^{t'} h(1)\, .
\end{equation}
The Asymptotic value of shocks   $h^\infty = \lim\limits_{t \rightarrow \infty} h(t)$ is finite if $\|\Lambda\| < 1$.  Gelfand theorem states that the spectral radius $\rho(\Lambda)=\max_{1\leq i \leq n}(\lambda_i)$  can for a square matrix always be written as
\begin{equation}
\rho(\Lambda) = \lim_{k \rightarrow \infty} \| \Lambda^k \|^{\frac{1}{k}}\, .
\end{equation}
If the spectral radius is smaller than 1 i.e. $|\lambda_{max}|<1$ Asymptotic value of financial shock converges towards 
\begin{equation}
h^\infty = (I - \Lambda)^{-1}h(1)\, \label{eq: h_inf}.
\end{equation}
In opposite case $|\lambda_{max}| \geq 1$, initial financial shock consumes the whole system.

It is, in principle, possible to  simulate the propagation of shocks such that when institution goes bankrupt  the matrix $\Lambda$ reduces its rank, but in the following we will focus our attention only on constant matrices $\Lambda$.

\section{Conditions on initial shocks for bankruptcy-free risk propagation}

The main scientific question of this paper is: Which initial shocks lead to systemic effects in financial system?
Clearly, the number of all possible configurations of initial shocks that could destroy financial system is huge and grows faster than the size of the system. In order to make our analysis as coherent as possible we focus on three different types of shocks. The first type of shocks we investigate is the uniform shock, which models a huge exogenous event in the system in which all the institutions loose the same fraction of assets. The second one is a shock to only one of the institutions in the system and we call it a localized shock. The last one we will investigate is a more general one mixing the previous two, that is a multi-parametric shock that affects all of the institutions with different intensities and will be measured through the associated hypervolume, as we will describe later.

\subsection{Uniform shock}

This type of shock represents one of the most commons types of macroeconomic shocks that can have different causes. It can be related to demand/supply shocks, changes in legislation that can affect financial markets, inflation {\em etc.} Financial systems are susceptible to outside conditions and are affected by different world events that often homogeneously impact similar firms. Financial systems are also built in such a way that they are relatively robust to these events, but if the shock is big enough uniform shocks are expected to lead to considerable problems for financial systems.

To model uniform shock we use equation \ref{eq: h_inf}.  Since $h(1)$ represents an initial value of shocks, we use the same value $h(1)$ for all the components of vector  i.e. for each $i$ $h_i(1)=\psi_u$. Than we can write

\begin{equation}
h(1) = \psi_u \begin{pmatrix}
1 \\ 1 \\ \vdots \\ 1
\end{pmatrix} = \psi_u \widetilde{h(1)}\, .
\end{equation}

If the uniform shock are used in equation \ref{eq: h_inf}
\begin{equation}
h^\infty = \psi_u (I - \Lambda)^{-1}  \widetilde{h(1)}\, .\label{eq:uni_eq}
\end{equation}

We denote as  $\Psi_u$ he maximal uniform shock that the system can absorb. To compute it, we need to maximize the equation \ref{eq:uni_eq}
 
 \begin{equation}
\Psi_u = \frac{1}{\max\limits_j \bigg[ \Big( (I - \Lambda)^{-1}  \widetilde{h(1)} \Big)_j  \bigg]} \, .
\end{equation}

The $\Psi_u$  depends only on the details of the network i.e. on the amounts of interbank assets and leverage of the banks. Larger values of $\Psi_u$ are associated to larger shocks the system can tolerate and therefore to a larger resilience of the system. Conversely, smaller values of $\Psi_u$ indicate systems more susceptible to the exogenous uniform risks. 

\subsection{Localised shock}

Compared to uniform shock, localized shock represents the opposite end of the spectrum of all possible shocks. Localised shock models financial difficulties related to a single financial institution, but the institution is systematically important i.e. its default can lead to instability of the whole financial system. Localized shocks had historically a significant influence on modern financial systems. For example, the default of Lehman Brothers in 2008 was a trigger event for the great recession. Lehman brothers was at a time fourth largest investment bank in the USA and its default propagated financial shocks to its counter-parties, contributing greatly to the greatest fall of Dow Jones index in history \cite{dow}.

Measure of systemic risk from uniform shock is developed in a similar way as  measures of uniform shock by using equation \ref{eq: h_inf}. The vector of initial shock $h_i(1)$ is the i-th unitary vector $\boldsymbol{e}_i$ multiplied by ${\psi_l}^i$
\begin{equation}
h(1)^i = {\psi_l}^i \begin{pmatrix}
0 \\  \vdots \\ 1 \\ \vdots \\ 0
\end{pmatrix} = {\psi_l}^i \boldsymbol{e}_i\, .
\end{equation}

Considering a network of $n$ banks, there are $n$ different initial conditions for localized shock, each of them representing the problem in the bank it represents. We are interested in maximal ${\psi_l}^i$ for which some element of $({h^\infty})^i$ becomes 1, 
\begin{equation}
\begin{split}
({h^\infty})^i_j &= {\psi_l}^i \big((I - \Lambda)^{-1}  \boldsymbol{e}_i \big)_j \\ &= {\psi_l}^i \sum_{k = 1}^n \big((I - \Lambda)^{-1}\big)_{jk}\delta_{ki} = {\psi_l}^i \big((I - \Lambda)^{-1}\big)_{ji}\, ,
\end{split}\label{eq: local Shock}
\end{equation}
where $\delta_{ki}$ is Kronecker delta.  We define \emph{maximal allowed localised shock} (MALS) as a value of local shock that Maximizes equation \ref{eq: local Shock}
\begin{equation}
{\Psi_l}^i = \frac{1}{\max\limits_j \Big[\big((I - \Lambda)^{-1}\big)_{ji} \Big]} \,.
\end{equation}

The ${\Psi_l}^i$ is determined only by the reduced adjacency matrix  $\Lambda$, and it depends only on one element of inverse matrix $(I - \Lambda)^{-1}$. One can interpret this measure as a measure
of the resilience of an individual institutions to starting
the financial bankruptcy cascade..

Generally instead to consider a single number we can focus on the components of  the vector $\Psi_l$ thus representing the overview of MALS for different elements of the network.
Important measure that we define from this consideration is a minimal component of maximally defined localised shock,
\begin{equation}
\Psi_l^m = \min\limits_i \Psi_l^i
\end{equation}
which signifies the largest allowed localized shock to any institution that can be sustained by the least resilient constituent of the system. 

\subsection{Heterogenous financial shock}

Uniform and localised financial shocks are two border cases that will lead to systemic risk in the financial system considered here. Both of these cases can be characterised through one parameter $\psi$ that can bu used to asses the riskiness of the system. Scalar measures like this are interesting to both academics and practitioners as they simplify evaluation of the system through just one measurable number. 

The suitability of these scalar measures, can be challenged in real situations in which a number of different shocks can simultaneously happen in the financial system. Indeed, in the general case of the financial system consisting of $n$ financial institutions, one can expect $n$ different parameters of financial shock, and would be interested in different combinations of initial shocks that still preserve the system stable. For that purpose we propose a measure related to such a shock, but which is still represented by a scalar quantity. This measure is again derived from the iterative evolution of cumulative capital loss \ref{eq: h_inf}, but we will observe this equation as a linear transformation of the allowed subspace $h^\infty$  

\begin{equation}
h(1) = (I - \Lambda)h^\infty\, .\label{eq: InitEvol}
\end{equation}

Taking in mind that the default of financial institution $i$ happens when $h^\infty_i = 1$, the total subspace of values for which there are no defaults in the system  ($0 \leq h^\infty_i < 1$) corresponds to $n$-hypercube, which is located at the origin of coordinate space defined with the axes $h^\infty_i$. Therefore, all the allowed cases in which no financial institution is damaged are located within the hypercube. 

To know which values of subspace of the parameters of initial shock $h(1)$ correspond to the system without defaults we transform the hypercube  from the space of asymptotic shock to the subspace of initial shock, using equation \ref{eq: InitEvol}.
 
The elements of the matrix $\Lambda$ are nonnegative, and the diagonal elements are always equal to zero, assuming that the institutions do not invest in themselves. This means that the most general matrix $I - \Lambda$ has ones on diagonal and all other elements are non positive i.e.

\begin{equation}
I - \Lambda = 
\begin{pmatrix}
1 & -\Lambda_{12} & \hdots & -\Lambda_{1n} \\
-\Lambda_{21} & 1 & & \vdots \\
\vdots & & \ddots &   \\
-\Lambda_{n1} & \hdots &  & 1
\end{pmatrix} \, .
\medskip
\end{equation}

As regards the linear transformation of hypercube this means that the subspace $h(1)$ will necessarily include values outside of the strictly positive $2^n$-tant, i.e. values of $h(1)_i$ for which there are no defaults will include values outside of realistic interval $[0,1 \rangle$. Those values should not be used since they represent the positive initial shock meaning increase of the institutional assets. Therefore the subspace of interest is an intersection   $V$ of allowed   $h(1)$ subspace and positive $2^n$-tant. 



Depending on the matrix $\Lambda$ the subspace of allowed initial shocks can be larger or smaller. In other words, if the network structure is such that the subspace of initial shocks $V$ is small, and we know that all the other combinations of initial shock lead to the default of at least one institution, than this network structure is of low resiliency. On the other hand if $V$ is of the size similar to the size of the subspace $h^\infty$, the probability that the initial combination of shocks will lead to default is small. 

Therefore, it makes sense to measure the risk of the system in terms of \emph{hypervolume} $\Psi_V$ of allowed subspace $V$. If the intersection with $2^n$-tant were not important, the computation of this measure would trivially be $\det (I - \Lambda)$. Unfortunately, depending on the sum of inter-banks investments, the geometry of the problem can be extremely complex even for a small number of vertices. 

In order to demonstrate the complexity and provide some intuition of analytical analysis we perform exact computations on a few small networks.

\section{Results}
\subsection{Network of $\mathbf{n = 2}$ vertices}

\begin{figure}[t]
  \centering
    \includegraphics[width=0.3\textwidth]{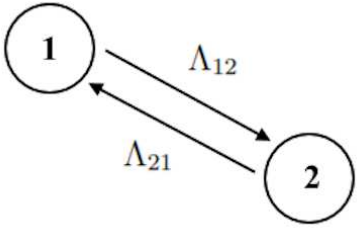}
 \caption{Network of $n = 2$ vertices with mutual exposures.}\label{Fig: 2 nodes}
\end{figure}

The simplest possible network of interest is the network of two institutions, creating $n = 2$ network as represented in Figure \ref{Fig: 2 nodes}. 

In that case the matrix of transformation is 
\begin{equation}
I - \Lambda = 
\begin{pmatrix}
1 & -\Lambda_{12} \\
-\Lambda_{21} & 1
\end{pmatrix} \, ,
\end{equation}

while its inverse is

\begin{equation}
(I - \Lambda)^{-1} = \frac{1}{1 - \Lambda_{12}\Lambda_{21}}
\begin{pmatrix}
1 & \Lambda_{12} \\
\Lambda_{21} & 1
\end{pmatrix} \, .
\end{equation}
We can easily compute measures of uniform and localised shock as

\begin{align}
\Psi_u &= \frac{1 - \Lambda_{12}\Lambda_{21}}{1 + \max \big[\Lambda_{12},\,\Lambda_{21}\big] }\, ,\\
\Psi_l &=  (1 - \Lambda_{12}\Lambda_{21}) 
\begin{pmatrix}
\min \big[ 1,\, \frac{1}{\Lambda_{21}} \big] \\
\min \big[ 1,\, \frac{1}{\Lambda_{12}} \big] 
\end{pmatrix} \, .
\end{align}
Hypervolume in this case corresponds to the surface of parallelogram which is computed through transformation and is confined in the first quadrant. The measure of heteorogenous shock is then
\begin{equation}
\Psi_V = 1 - \frac{1}{2}\big(\Lambda_{12}(1 - \Lambda_{21}^2) + \Lambda_{21}(1 - \Lambda_{12}^2)\big) \, .
\end{equation}
Allowed intervals of initial shock which do not produce default in the system can be computed using relations  $0 \leq h^\infty_i < 1,\, i \in \{1,2 \}$ which leads to

\begin{align}
\begin{split}
0 \leq \frac{h(1)_1 + \Lambda_{21}h(1)_2}{1 - \Lambda_{12}\Lambda_{21}} < 1 \\
0 \leq \frac{\Lambda_{12}h(1)_1 + h(1)_2}{1 - \Lambda_{12}\Lambda_{21}} < 1
\end{split} \,.
\end{align}

\subsection{Network of $\mathbf{n = 3}$ vertices}

\begin{figure}[t]
     \centering
         \includegraphics[width=0.3\textwidth]{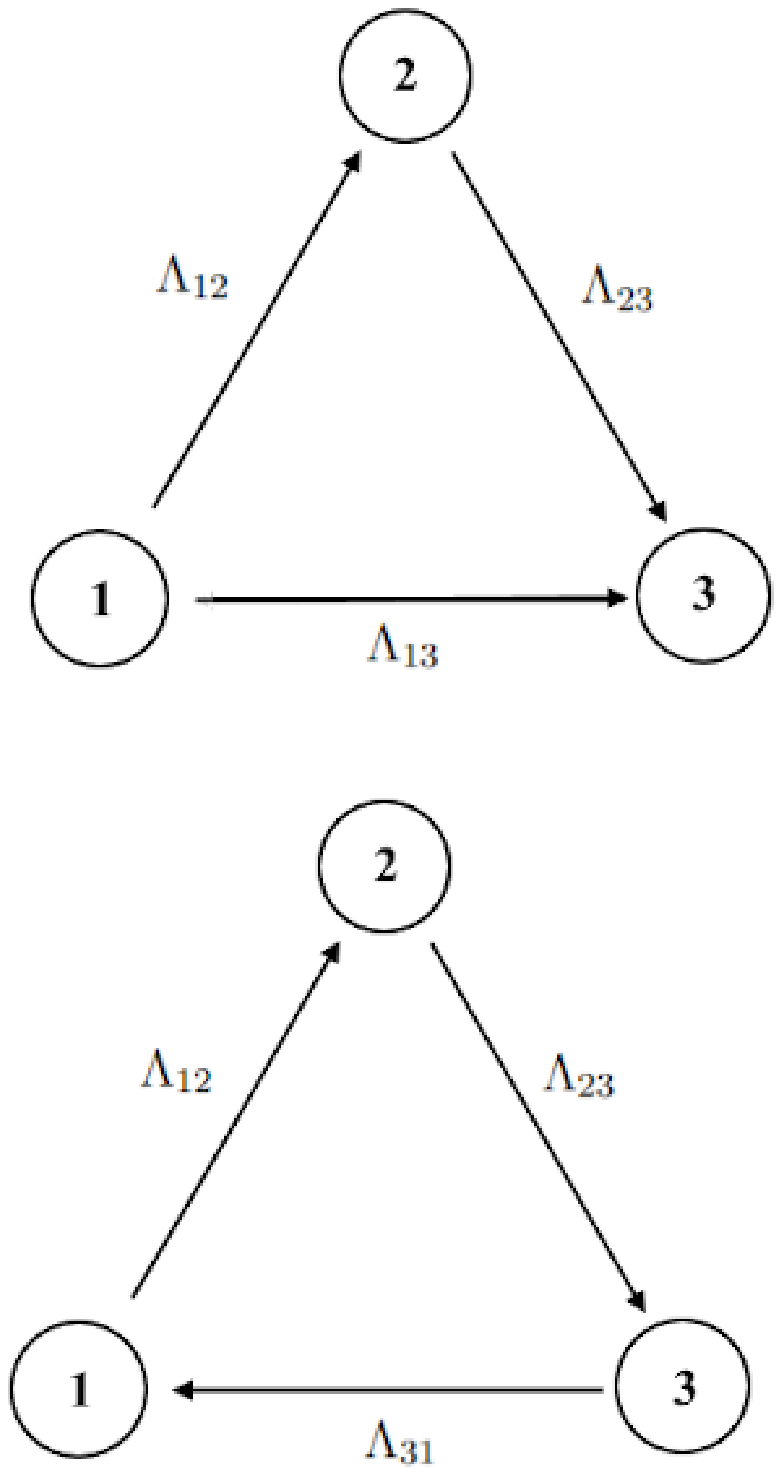}
    
         \caption{In the upper panel is Uppertriangular and in the lower panel is a cyclical configuration of network with $n = 3$ vertices.}\label{Fig: 3nodes}
\end{figure}

In the case of $n=3$ there are two independent cases for which one can obtain all the other configurations through cyclical change of indices of $\Lambda_{ij}$. We omit the cases with two edges as they are not that interesting. Two possible configurations of three vertices and three edges are presented in Figure \ref{Fig: 3nodes}.

The set of parameters that given the evolution of financial shocks is represented by matrix $I - \Lambda$ and configuration with $n = 3$, will according to its matrix type will be called uppertriangular (index  \textit{UT}) and cyclical (index \textit{cyc}). Matrices of evolution for both cases are given by 

\begin{align}
I - \Lambda_{UT} &= 
\begin{pmatrix}
1 & -\Lambda_{12} & -\Lambda_{13} \\
0 & 1   & -\Lambda_{23}  \\
0 & 0 & 1
\end{pmatrix} \, , \\
\smallskip
I - \Lambda_{cyc}& = 
\begin{pmatrix}
1 & -\Lambda_{12} & 0 \\
0 & 1   & -\Lambda_{23}  \\
 -\Lambda_{31} & 0 & 1
\end{pmatrix} \, .
\end{align}

 Inverses of $I - \Lambda$ matrices, are needed for computation of measures of risk and are given by

\begin{align}
(I - \Lambda_{UT})^{-1} &= 
\begin{pmatrix}
1 & \Lambda_{12} & \Lambda_{12}\Lambda_{23} +\Lambda_{13} \\
0 & 1   & \Lambda_{23}  \\
0 & 0 & 1
\end{pmatrix} \, , \\
\medskip
(I - \Lambda_{cyc})^{-1}& = \frac{\begin{pmatrix}
1 & \Lambda_{12} & \Lambda_{12}\Lambda_{23} \\
\Lambda_{23}\Lambda_{31} & 1   & \Lambda_{23}  \\
 \Lambda_{31} & \Lambda_{31}\Lambda_{12} & 1
\end{pmatrix}}{1 - \Lambda_{12}\Lambda_{23}\Lambda_{31}}
 \, .
\end{align}

\begin{figure}[t]
  \centering
    \includegraphics[width=0.48\textwidth]{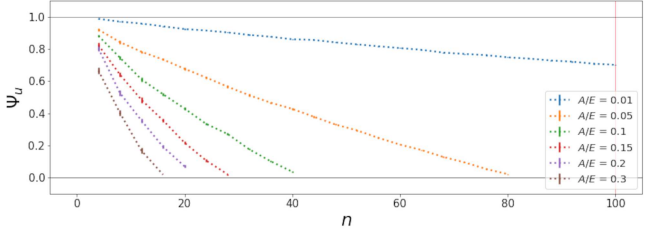}
 \caption{On y-axis is Maximal allowed uniform shock on CASD network $\Psi_u$, on x-axis is the size of the system $n$, $n \in [5, 100]$. Investigated parameters $A/E$ are listed in the legend.}\label{Fig: Psi_u CASD on n}
\end{figure}

Similar to $n = 2$ case, maximal uniform and localized shocks are for uppertriangular case
\begin{align}
\Psi_u^{UT} &= \frac{1}{1 + \max \big[\Lambda_{13} + \Lambda_{12}( 1 + \Lambda_{23}),\, \Lambda_{23} \big]}\, ,\\[10pt]
\Psi_l^{UT} &=   
\begin{pmatrix}
1 \\
\min \big[ 1,\, \frac{1}{\Lambda_{12}} \big] \\
\min \big[ 1,\, \frac{1}{\Lambda_{23}},\, \frac{1}{\Lambda_{13} + \Lambda_{12}\Lambda_{23}} \big] 
\end{pmatrix} \, ,
\end{align}
and for the cyclical configuration
\begin{align}
\Psi_u^{cyc} &= \frac{1 - \Lambda_{12}\Lambda_{23}\Lambda_{31}}{1 + \max \big[   \Lambda_{ij}(1 +\Lambda_{jk})  \big] }\, ,\\[10pt]
\Psi_l^{cyc} &=  (1 - \Lambda_{12}\Lambda_{23}\Lambda_{31}) 
\begin{pmatrix}
\min \big[ 1,\, \frac{1}{\Lambda_{31}},\, \frac{1}{\Lambda_{23}\Lambda_{31}} \big] \\
\min \big[ 1,\, \frac{1}{\Lambda_{12}},\, \frac{1}{\Lambda_{31}\Lambda_{12}} \big] \\
\min \big[ 1,\, \frac{1}{\Lambda_{23}},\, \frac{1}{\Lambda_{12}\Lambda_{23}} \big] 
\end{pmatrix} \, ,
\end{align}
where in the first equation one substitutes corresponding indices $i,j,k$ with those for with the maximal value in the square bracket. 

For studied configurations of the $n=3$ network we can again obtain intervals of allowed initial shocks using conditions $0 \leq h^\infty_i < 1 ,\, i \in \{1,2,3 \}$. As in the previous case the solution is a system of equations for uppertriangular configuration
\begin{align}
\begin{split}
0 \leq  h(1)_1 + \Lambda_{12}h(1)_2 &+ (\Lambda_{13} + \Lambda_{12}\Lambda_{23})h(1)_3 < 1 \\[5pt]
0 \leq  h(1)_2 &+ \Lambda_{23}h(1)_3 < 1 
\end{split} \,,
\end{align}
and for the cyclical configuration
\begin{align}
\begin{split}
0 \leq \frac{h(1)_1 + \Lambda_{12}h(1)_2 + \Lambda_{12}\Lambda_{23}h(1)_3}{1 - \Lambda_{12}\Lambda_{23}\Lambda_{31}} < 1 \\[5pt]
0 \leq \frac{\Lambda_{23}\Lambda_{31}h(1)_1 + h(1)_2 + \Lambda_{23}h(1)_3}{1 - \Lambda_{12}\Lambda_{23}\Lambda_{31}} < 1 \\[5pt]
0 \leq \frac{\Lambda_{31}h(1)_1 + \Lambda_{31}\Lambda_{12}h(1)_2 + h(1)_3}{1 - \Lambda_{12}\Lambda_{23}\Lambda_{31}} < 1
\end{split} \,.
\end{align}

\begin{figure}[t]
\includegraphics[width=0.48\textwidth]{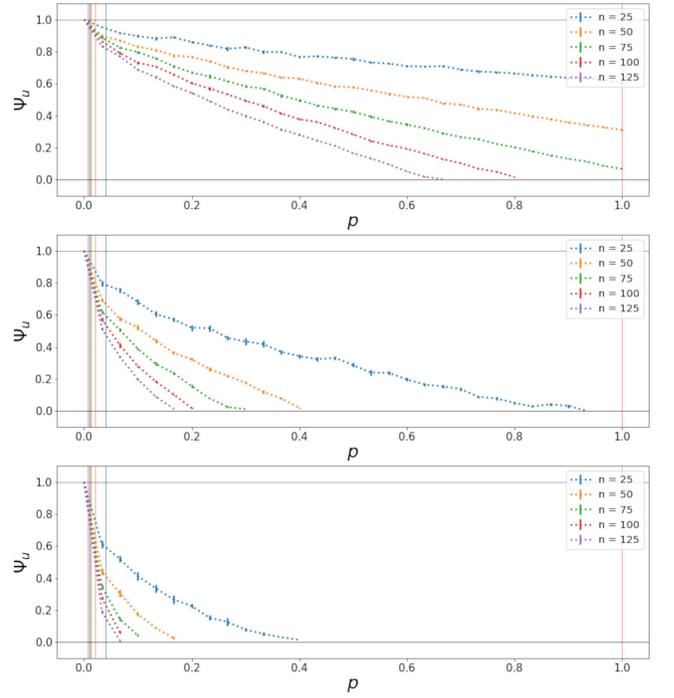}
\caption{Maximal allowed uniform shock on incomplete random network. On y-axis is $\Psi_u$, and on x-axis parameter of connectivity $p$. Sizes of networks are in legend and are $n \in \{25, 50, 75, 100, 125 \}$, while vertical lines correspond to  $p_c$ for different values of  $n$. The top, middle and bottom subfigures correspond to parameters $A/E$ equal to $0.05$, $0.2$ i $0.5$, respectively. Values of $p_c$ are indicated by vertical colored lines.}\label{Fig: Psi_u on p}
\end{figure}

Hypervolume measure of heterogeneous shock $\Psi_V$ on the network of $n = 3$ vertices corresponds to the volume of the parallelepiped in $h(1)$ space obtained through  $I - \Lambda$ transform, that is inside the first octant. Volume $\Psi_V$ can be computed geometrically, by cutting the parellelepiped into prisms and pyramids of known volume, but already in 3d the number of such elements becomes large, as well as the number of different geometries depending on elements $\Lambda_{ij}$. For example the equation for hypervolume associated with uppertriangular configuration is
\begin{equation}
\begin{split}
\Psi_V^{UT} =\,  &\frac{1}{2}\Lambda_{12}\big(1 - \Lambda_{23}\big)^2 + \frac{1}{2}\big(\Lambda_{13} + \Lambda_{12}\Lambda_{23}\big)\big(1 - \frac{2}{3}\Lambda_{23}\big)\\ &+ \frac{1}{6}\Lambda_{13}\Lambda_{23}  + \big(1 - \Lambda_{23} + \frac{1}{2}\Lambda_{23}\big)\big(1 - \Lambda_{12} - \Lambda_{13}\big)\,;\\ &\quad \Lambda_{23} < 1\, ,\, \Lambda_{12} + \Lambda_{13} < 1\, .
\end{split}
\end{equation}

\begin{figure}[t]
     \centering
         \includegraphics[width=0.48\textwidth]{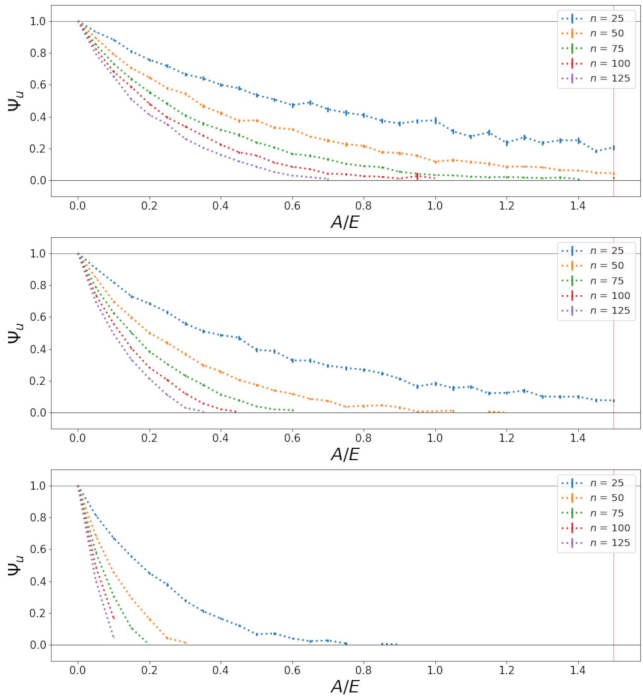}
        \caption{Maximally allowed uniform shock of incomplete random network depending on the parameter $A/E$. On y-axis is a measure $\Psi_u$, and on x-axis is  $A/E \in \langle 0, 1.5]$. Sizes of networks are in legend and they take values $n \in \{25, 50, 75, 100, 125 \}$. Top, middle and bottom subfigure correspond to parameters $p$ equal to $0.05$, $0.1$ and $0.3$, respectively.}
        \label{Incomplete On A/C}
\end{figure}

\subsection{Multiple vertices}

     
  
    
     

Clearly with increase of the size of the system analytical methods to compute this measures become complicated and for this reason heterogenous shock $\Psi_V$ is computed using MonteCarlo simulations. 

In $h^\infty$ space $M$ random allowed vectors with components $0 \leq h^\infty_i <1$ from uniform distribution are generated. The hypervolume of allowed $h^\infty$ subspace is always one.  The created vectors are then transformed to $h(1)$ space, using equation \ref{eq: InitEvol}. With $M'$ we designate the number of points within the intersection of polytope and positive $2^n$-tant. Since the number of generated points is proportional to the hypervolume $\det(I - \Lambda)$ of the polytop, the volume $\Psi_V$ of allowed initial conditions is estimated through
\begin{equation}
\Psi_V = \frac{M'}{M}\det(I - \Lambda)\, .
\end{equation}

In order to estimate risk measures in more complicated architectures we shall use a complete asymmetric simple digraph (CASD) and our variant of random network model. The reason to use a complete asymmetric simple digraph is that a number of financial networks are very densely connected and the complete network represents a limit of large density as well as the most diversified investment pattern - which is in general accepted to be a pattern that reduces individual risk but may increase systemic risk \cite{battiston2012liaisons}.  

The CASD is produced generating the matrix  $\Lambda$. Algorihm used to generate CASD is explained in the following steps: 

\begin{enumerate}
\item We choose $n$ vertices.
\item We assume that the equity of all constituents are equal i.e.  $E_i = E$, and investments among the banks $A_{ij}$ are chosen randomly from the uniform distribution on interval  $[0,A]$.
\item for each pair of indices $i,j<n;\, i<j$ a random number drawn from uniform distribution  $p'\in[0,1]$  is generated. If $p'<0.5$ to the element $A_{ij}$  we assign a random value, and in the opposite case the same value is assigned to the element  $A_{ji}$.
\item Matrix elements $A_{ij}$ are divided with the constant equity $E$ leading to reduced adjacency matrix $\Lambda$.
\item We check if the spectral radius of matrix $\Lambda$ is smaller than 1, if not, we go back to the step 1.
\end{enumerate}

Except by parameters $A$ and $E$, incomplete random network is also described by additional parameter $p$ giving the probability that the two randomly chosen vertices are connected with a directional edge. In order to produce this network in above algorithm after the second step we introduce intermediate step in which a parameter 
$p \in [0,1]$ is chosen and for each pair of vertices  $i,j<n;\, i<j$ we generate a random number $p'' \in [0,1]$. If $p'' < p$, the vertices will be connected, and otherwise they won't be connected. After this step, steps 3-5 are repeated.

\begin{figure}[t]
  \centering
    \includegraphics[width=0.48\textwidth]{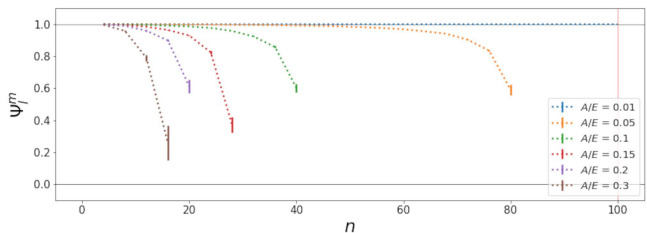}
 \caption{Maximally allowed localised shock on CASD depending on the number of vertices.  On y-axis is $\Psi_l^m$ , while on x-axis is the size of the network $n= \in [5, 100]$. Parameter $A/E$ is shown in legend and it takes values $A/E \in \{0.01, 0.05, 0.1, 0.15, 0.2, 0.3 \}$. }
\end{figure}

\subsubsection*{Uniform shock $\Psi_u$}

Measure of uniform shock is first computed for the case of CASD. Main interest is to understand dependence of uniform shock $\Psi_u$ on different network parameters. Simulation is performed on ensemble of $N = 100$ networks. We first investigate how uniform shock depends on the size of the network. Results are presented in Figure \ref{Fig: Psi_u CASD on n}. Error bars represent standard deviation of simulation results $\sigma_{\bar{\Psi}_u}$, as will be the case in all other simulations. The first interesting observation is that the size of the system systematically reduces $\Psi_u$ for all the choices of other parameters, which in essence signals that larger systems are always more risky than smaller once with respect to uniform shocks. Parameter $A/E$ is related to speed with which $\Psi_u$ tends to zero.

The computation of $\Psi_u$ is continued on the ensemble of incomplete random networks characterised by parameters  $n$, $p$ and $A/E$. We first check dependence of the measure $\Psi_u$ on parameter $p$. Since for the values of parameter smaller than  $p_c \approx \frac{1}{n}$, networks do not posses giant component we present only the results for $p> p_c$. Dependence of $\Psi_u$ on $p$ is shown in Figure \ref{Fig: Psi_u on p}. It is clear that the increase of parameters $p$ and $n$ reduces the allowed maximal uniform shock systematically for all the observed parameter values. This means that both diversification and size of the system negatively affect maximal uniform shocks that the institutions can accommodate previous to observation of defaults in the system. 

Furthermore in Figure \ref{Incomplete On A/C} for completeness we show how value of $\Psi_u$ depends on $A/E$, although it is intuitively clear that in will get smaller with the increase of $A/E$.
\begin{figure}[t]
     \centering

    \includegraphics[width=0.48\textwidth]{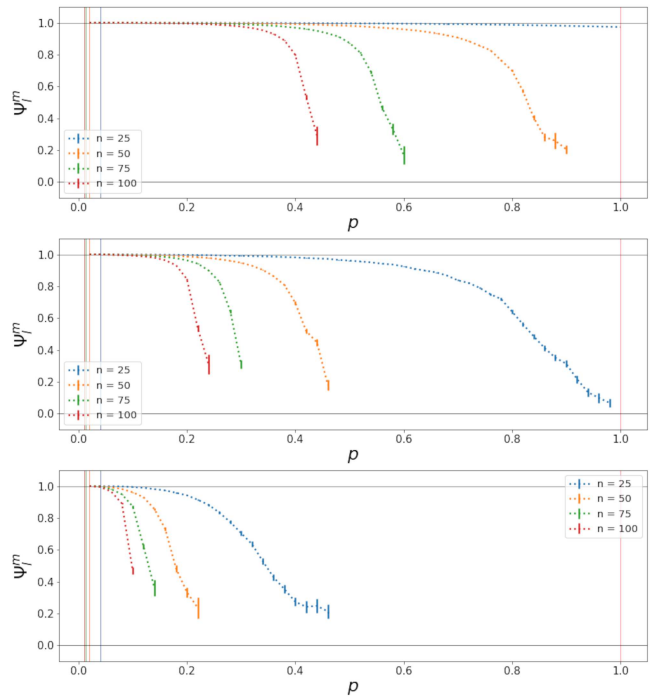}

        \caption{$\Psi_l^m$ on random directed network depending on parameter $p$. On y-axis is $\Psi_l^m$, while on x-axis is $p \in \langle 0, 1]$. Network sizes $n \in \{25, 50, 75, 100\}$, are presented in the legend, while vertical lines correspond to critical values of $p_c$. Top, middle and bottom subfigure correspond to parameters  $A/E$ equal to $0.1$, $0.2$ and $0.5$, respectively.}\label{fig: RN PSIlm on p}
\end{figure}

\subsubsection*{Measure of localised shock $\Psi_l$}

All the analysis related to uniform financial shock, we repeat for the localised shock. We focus our attention to the minimal component $\Psi_l^m$ of maximally allowed vector of localized shock $\Psi_l$. 

Compared to the case of uniform shock there is a more sever influence of parameter $A/E$ on the decline of $\Psi_l^m$ depending on the number of vertices $n$. Corresponding curves are now concave as opposed to convex curves present in uniform shock. After a period of relatively stable behavior (that is shortened with increase of $A/E$ there is a sudden drop in minimal value of maximally allowed local shock. The measure $\Psi_l^m$ further declines to the value after which generated matrices $\Lambda$ have spectral radius larger then one. 

\begin{figure}[t]
     \centering
    
    \includegraphics[width=0.48\textwidth]{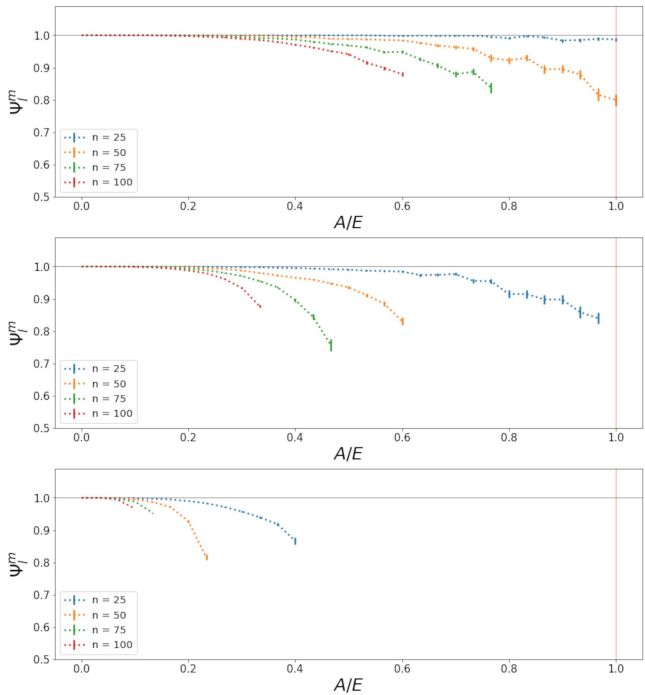}
     
        \caption{$\Psi_l^m$ on random directed network depending on parameter  $A/E$. On y-axis is $\Psi_l^m$, while on x-axis is  $A/E \in \langle 0, 1]$. Network sizes $n \in \{25, 50, 75, 100\}$, are presented in the legend. Top, middle and bottom subfigure correspond to parameters  $p$ equal to $0.05$, $0.1$ and $0.3$, respectively.}\label{fig: RN PSIlm on A/C}
\end{figure}

We again study the risk on random network varying parameters $p$ and $A/E$. In this case we used ensembles of $N=100$ networks to have satisfactory stability of the results. Simulations are depicted in Figures \ref{fig: RN PSIlm on p} and \ref{fig: RN PSIlm on A/C}.

In both cases as $n$, $p$ and $A/E$ increase the $\Psi_l^m$ decreases monotonically. Compared to incomplete random network this decrease is steeper for $\Psi_l^m$, than for uniform shock $\Psi_u$, indicating that this parameters more strongly influence localized shocks compared to uniform once. 

An important addition to this analysis came from graph theory \cite{kwapisz1996spectral}, which gives upper bound of spectral radius = $\max \{ \sqrt{s_is_o}\}$, where $s_i$ and $s_o$ are strengths of vertices. In our case this values are approximately of the order $A/E$, and are a leading contribution to this sudden collapse of the $\Psi_l^m$. 


\begin{figure}[t!]
     \centering
         \includegraphics[width=0.48\textwidth]{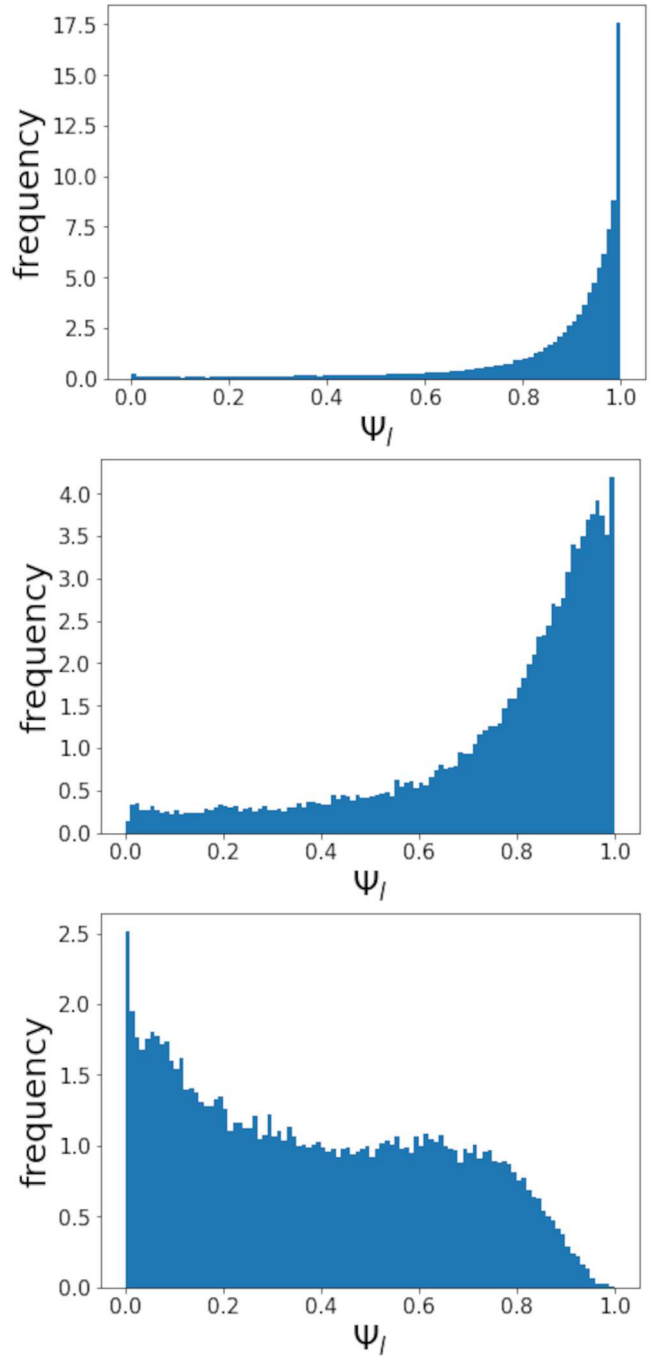}
        
        \caption{Histograms of vector components of localised shock $\Psi_l$ on ensemble of $N = 1000$ random networks. The x-axis is binned on 100 bins and the values are normalised to make surface equal to 1. The parameters used for the top panel are $n = 75, p = 0.1, A/E = 0.5$. The parameters used for the middle panel are $n = 50, p = 0.2, A/E = 0.4$. The parameters used for the bottom panel are $n = 50, p = 0.4, A/E = 0.23$. }
        \label{fig: Hist PSIl of vector}
\end{figure}

Up to now, we have considered the measure of localised shock $\Psi_l^m$, but if we want to understand more details of the distribution of risks in the network, we can take into consideration all the components of vector $\Psi_l$, i.e. a maximal shock with origin in any of the vertices in the network. 

For example, we have chosen representative parameters to represent three different states of the system. In the top panel of the Figure \ref{fig: Hist PSIl of vector} most of the values are located around value 1, signifying overall stability of the network. In the middle panel of the gigure \ref{fig: Hist PSIl of vector} the distribution of values exhibit move to the smaller values, indicating more stressed system. In the bottom panel of the Figure \ref{fig: Hist PSIl of vector}, values are starting to cluster around $\Psi_l\approx 0$ signifying system close to falling a part. Such a figure can give an important information of the risk of the system at a glance.

\subsubsection*{Measure of multi-parametric shock $\Psi_V$}

In the end we consider the behavior of hypervolume associated shock $\Psi_V$ dependence on network parameters. For the approximation of $n$-dimensional hypervolume, we use previously described Monte Carlo method. 

\begin{figure}[t]
  \centering
    \includegraphics[width=0.48\textwidth]{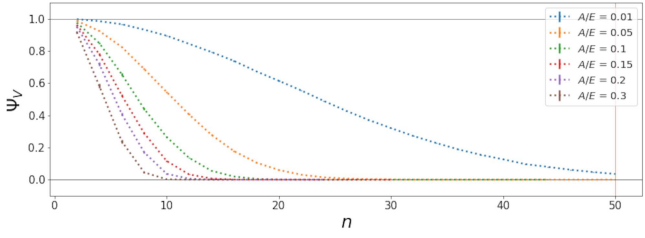}
 \caption{Hypervolume measure $\Psi_V$ evaluated on the CASD network depending on the number of vertices. Computation is obtained through $N=50$ realization for each of parameters $n \in [2, 50]$ and for parameters $A/E \in \{0.01, 0.05, 0.1, 0.15, 0.2, 0.3 \}$. We have used $M=10^4$ points to evaluate hypervolume.  }\label{fig: PsiV on CASD}
\end{figure}

Comparing results from Figure \ref{fig: PsiV on CASD}, with previous Figures one can observe that the measure $\Psi_V$ is the most restrictive regarding the speed of decline with respect to $n$. This means that it evaluates the parameters of the system as more risky then other measures. 

\begin{figure}[t!]
     \centering
         \includegraphics[width=0.48\textwidth]{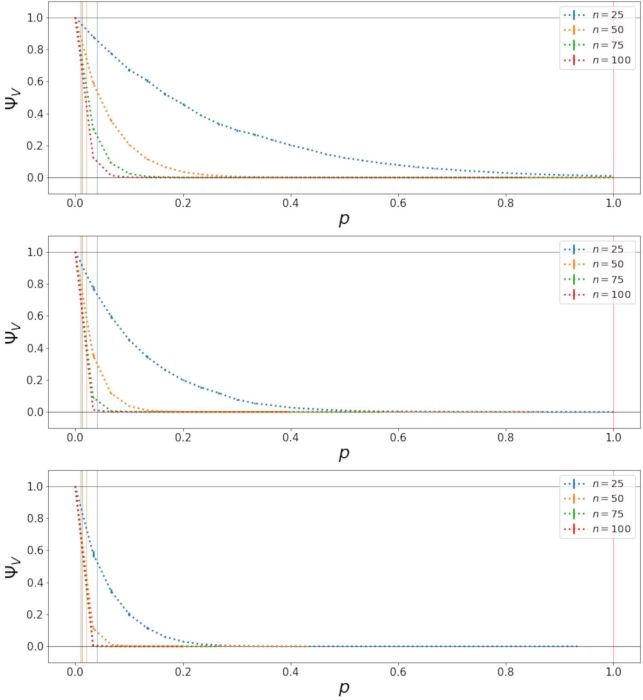}

        \caption{Measure of hypervolume $\Psi_V$ on random network depending on parameter $p$. Computations are evaluated on ensemble of $N=50$ realizations for  $p \in \langle 0, 1]$ , and $n \in \{25, 50, 75, 100\}$. Vertical lines correspond to critical values of $p_c$ for percolation on random networks. Figures (a), (b) and (c) correspond to  parameters $A/E$ equal to $0.05$, $0.1$ and $0.2$. Monte Carlo is algorithm used $M = 10^4$ points.}
        \label{fig: PsiV Incomplete p}
\end{figure}

\begin{figure}[t!]
     \centering
\includegraphics[width=0.48\textwidth]{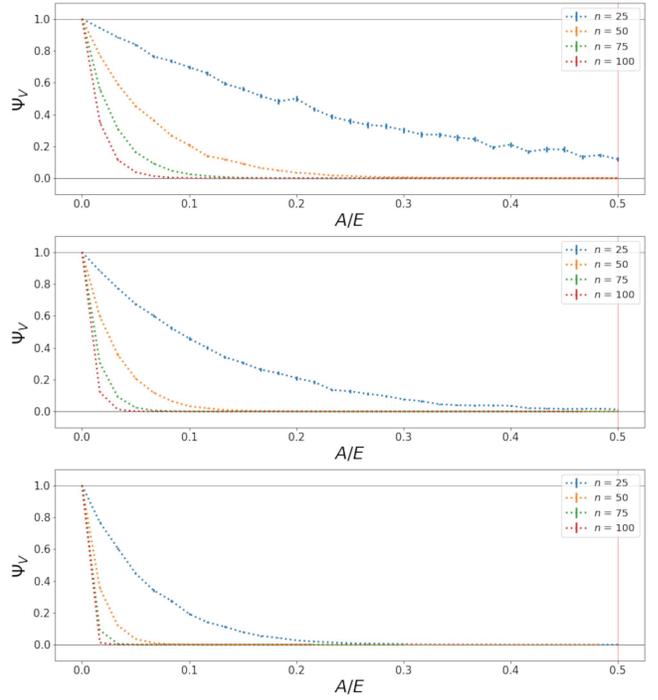}

        \caption{Hypervolume measure $\Psi_V$ on incomplete random network depending on parameter $A/E$. Computation is carried out on ensemble of $N=50$ networks for $A/E \in \langle 0, 1]$, and $n \in \{25, 50, 75, 100 \}$. Figures correspond to parmeters $p$ equal to $0.05$, $0.1$, and $0.2$, respectively. Hypervolume is evaluated using $M=10^4$ points.}
        \label{fig: PsiV Incomplete A/C}
\end{figure}

On incompletely connected network the tendency of steeper decline of the $\Psi_V$  is again visible in Figure \ref{fig: PsiV on CASD}. This results hold for increase of any parameter $n$, $p$ or $A/E$ as can additionally be seen in Figures \ref{fig: PsiV Incomplete p} and \ref{fig: PsiV Incomplete A/C}. We believe that the reason for this behavior is related to the very nature of this measure. Hypervolume measure $\Psi_V$ contains a marge larger combination of initial shocks that will lead to some bankruptcy, than in the cases of measures $\Psi_u$ and $\Psi_l$. Here it is important to stress that hypervolume measure is inherently different from other proposed measures as it does not measure allowed shock but the relative size of allowed parameter space for which shocks are mitigated. Since the hypervolume is of dimension equal to the number of institutions in the network D=n, and the first 2 measures are 1D and actually represent measures on a vectors either spanning this hypervolume (localised shocks) or representing a diagonal (uniform shocks) one could naively expect that the region safe from heterogenous shock scales as $\sim h^n$, thus making its ratio to the complete volume very small for large system size $n$.

It is also important to stress that the sampling of higherdimensional polythopes in computationally expensive task if the goal is to minimise relative errors. In that sense $M=10^4$ points would not be enough, however (i) the goal of this research was to demonstrate the principle and (ii) we do not evaluate one polythope but we compute on ensemble of polythopes which should in principal reduce relative error of algorithm.



The largest relative errors are in points where exact hypervolume is finite but very small and the Monte Carlo measure hypervolume is zero. On the other hand the relative error is much smaller for the more common parameter choices. This is important from practical reasons, since in financial system a shock of small value is a common daily occurrence, while Monte Carlo method gives better estimates for more realistic values of shock. In real systems one can use more advanced algorithms for approximation of hypervolume like \textit{VolEsti}~\cite{chalkis2020volesti}. Another possible extension is using the historic measures of individual shocks to sample points proportionally to expected probability of occurrence, which would better describe border cases. 

\section{Conclusions}

This research considered three proposed measures for systemic risk in finance. We have used uniform shock measure $\Psi_u$ to estimate the maximal simultaneous fall of capital in the system that can happen in all institutions without an of them failing. Maximal localized shock $\Psi_l$ is a vector of largest shocks that an individual institution can suffer without bankrupting itself or some other institution in the system. The hypervolume measure $\Psi_V$ is a generalized measure that takes into account all the intervals of initial shocks and can be evaluated only approximately. 

We have evaluated the equations of measures analytically for small networks $n \leq 3$, to get the intuition on importance of different parameters. 

Furthermore, we have used a large ensemble of random networks to evaluate proposed measures on the more realistic models of financial system. Here we had some important ovbservations: 

First, among the presented measures, the measure of uniform shock is decreasing most slowly with increase of system parameters, signifying that the systems are most resilient against broad uniform outside shocks to the system. For all the range of parameters the functions seem to be concave.

Second, the measure of localised shock  decreases slowly with increase of  parameters, until the region in which the system abruptly becomes very susceptible to the shock of one institution. For all the ranges of parameters we have found that the functions seem to be convex. 

Third, the hypervolume measure that estimates the heterogenous shocks is the most conservative of these measures, as it  decreases steepest, among the studied measures, with the increase of parameters. It is also the measure that mixes concave with convex regions as a function of model parameters.  

We believe that this methodology can be of interest to regulators as it gives information about how initial shocks affect other constituents of the financial system, and can point to institutions whose bad behavior can seriously affect other well behaved institutions. We hope that we will have a chance to evaluate this measures on real financial data in the future in order to better understand their behavior in a more realistic settings.

\begin{acknowledgments}
HS and VZ had their research supported by the European Regional Development Fund under the Grant KK.01.1.1.01.0009 (DATACROSS).
GC acknowledges support from EU Project "HumanE-AI-Net", no. 952026.
VZ wish to acknowledge the support of the Croatian Science Foundation (HrZZ) Projects No. IP-2019-4-3321 and acknowledges partial support form QuantiXLie Centre of Excellence, a Project co-financed by the Croatian Government and European Union through the European Regional Development Fund - the Competitiveness and Cohesion Operational Program (Grant KK.01.1.1.01.0004, element leader N.P.).
\end{acknowledgments}

\pagebreak

\bibliography{reference}

\begin{thebibliography}{23}%
\makeatletter
\providecommand \@ifxundefined [1]{%
 \@ifx{#1\undefined}
}%
\providecommand \@ifnum [1]{%
 \ifnum #1\expandafter \@firstoftwo
 \else \expandafter \@secondoftwo
 \fi
}%
\providecommand \@ifx [1]{%
 \ifx #1\expandafter \@firstoftwo
 \else \expandafter \@secondoftwo
 \fi
}%
\providecommand \natexlab [1]{#1}%
\providecommand \enquote  [1]{``#1''}%
\providecommand \bibnamefont  [1]{#1}%
\providecommand \bibfnamefont [1]{#1}%
\providecommand \citenamefont [1]{#1}%
\providecommand \href@noop [0]{\@secondoftwo}%
\providecommand \href [0]{\begingroup \@sanitize@url \@href}%
\providecommand \@href[1]{\@@startlink{#1}\@@href}%
\providecommand \@@href[1]{\endgroup#1\@@endlink}%
\providecommand \@sanitize@url [0]{\catcode `\\12\catcode `\$12\catcode
  `\&12\catcode `\#12\catcode `\^12\catcode `\_12\catcode `\%12\relax}%
\providecommand \@@startlink[1]{}%
\providecommand \@@endlink[0]{}%
\providecommand \url  [0]{\begingroup\@sanitize@url \@url }%
\providecommand \@url [1]{\endgroup\@href {#1}{\urlprefix }}%
\providecommand \urlprefix  [0]{URL }%
\providecommand \Eprint [0]{\href }%
\providecommand \doibase [0]{http://dx.doi.org/}%
\providecommand \selectlanguage [0]{\@gobble}%
\providecommand \bibinfo  [0]{\@secondoftwo}%
\providecommand \bibfield  [0]{\@secondoftwo}%
\providecommand \translation [1]{[#1]}%
\providecommand \BibitemOpen [0]{}%
\providecommand \bibitemStop [0]{}%
\providecommand \bibitemNoStop [0]{.\EOS\space}%
\providecommand \EOS [0]{\spacefactor3000\relax}%
\providecommand \BibitemShut  [1]{\csname bibitem#1\endcsname}%
\let\auto@bib@innerbib\@empty
\bibitem [{\citenamefont {Newman}\ \emph {et~al.}(2006)\citenamefont {Newman},
  \citenamefont {Barab{\'a}si},\ and\ \citenamefont {Watts}}]{NewmanNetworks}%
  \BibitemOpen
  \bibfield  {author} {\bibinfo {author} {\bibfnamefont {M.~E.}\ \bibnamefont
  {Newman}}, \bibinfo {author} {\bibfnamefont {A.-L.~E.}\ \bibnamefont
  {Barab{\'a}si}}, \ and\ \bibinfo {author} {\bibfnamefont {D.~J.}\
  \bibnamefont {Watts}},\ }\href@noop {} {\emph {\bibinfo {title} {The
  structure and dynamics of networks.}}}\ (\bibinfo  {publisher} {Princeton
  university press},\ \bibinfo {year} {2006})\BibitemShut {NoStop}%
\bibitem [{\citenamefont {Chinazzi}\ and\ \citenamefont
  {Fagiolo}(2015)}]{chinazzi2015systemic}%
  \BibitemOpen
  \bibfield  {author} {\bibinfo {author} {\bibfnamefont {M.}~\bibnamefont
  {Chinazzi}}\ and\ \bibinfo {author} {\bibfnamefont {G.}~\bibnamefont
  {Fagiolo}},\ }\href@noop {} {\emph {\bibinfo {title} {Systemic risk,
  contagion, and financial networks: A survey}}}\ (\bibinfo  {publisher}
  {SSRN},\ \bibinfo {year} {2015})\BibitemShut {NoStop}%
\bibitem [{\citenamefont {Gai}\ and\ \citenamefont
  {Kapadia}(2019)}]{gai2019networks}%
  \BibitemOpen
  \bibfield  {author} {\bibinfo {author} {\bibfnamefont {P.}~\bibnamefont
  {Gai}}\ and\ \bibinfo {author} {\bibfnamefont {S.}~\bibnamefont {Kapadia}},\
  }\href@noop {} {\bibfield  {journal} {\bibinfo  {journal} {Oxford Review of
  Economic Policy}\ }\textbf {\bibinfo {volume} {35}},\ \bibinfo {pages} {586}
  (\bibinfo {year} {2019})}\BibitemShut {NoStop}%
\bibitem [{\citenamefont {Bardoscia}\ \emph {et~al.}(2021)\citenamefont
  {Bardoscia}, \citenamefont {Barucca}, \citenamefont {Battiston},
  \citenamefont {Caccioli}, \citenamefont {Cimini}, \citenamefont
  {Garlaschelli}, \citenamefont {Saracco}, \citenamefont {Squartini},\ and\
  \citenamefont {Caldarelli}}]{bardoscia2021physics}%
  \BibitemOpen
  \bibfield  {author} {\bibinfo {author} {\bibfnamefont {M.}~\bibnamefont
  {Bardoscia}}, \bibinfo {author} {\bibfnamefont {P.}~\bibnamefont {Barucca}},
  \bibinfo {author} {\bibfnamefont {S.}~\bibnamefont {Battiston}}, \bibinfo
  {author} {\bibfnamefont {F.}~\bibnamefont {Caccioli}}, \bibinfo {author}
  {\bibfnamefont {G.}~\bibnamefont {Cimini}}, \bibinfo {author} {\bibfnamefont
  {D.}~\bibnamefont {Garlaschelli}}, \bibinfo {author} {\bibfnamefont
  {F.}~\bibnamefont {Saracco}}, \bibinfo {author} {\bibfnamefont
  {T.}~\bibnamefont {Squartini}}, \ and\ \bibinfo {author} {\bibfnamefont
  {G.}~\bibnamefont {Caldarelli}},\ }\href@noop {} {\bibfield  {journal}
  {\bibinfo  {journal} {Nature Reviews Physics}\ }\textbf {\bibinfo {volume}
  {3}},\ \bibinfo {pages} {490} (\bibinfo {year} {2021})}\BibitemShut {NoStop}%
\bibitem [{\citenamefont {Battiston}\ \emph {et~al.}(2016)\citenamefont
  {Battiston}, \citenamefont {Farmer}, \citenamefont {Flache}, \citenamefont
  {Garlaschelli}, \citenamefont {Haldane}, \citenamefont {Heesterbeek},
  \citenamefont {Hommes}, \citenamefont {Jaeger}, \citenamefont {May},\ and\
  \citenamefont {Scheffer}}]{battiston2016complexity}%
  \BibitemOpen
  \bibfield  {author} {\bibinfo {author} {\bibfnamefont {S.}~\bibnamefont
  {Battiston}}, \bibinfo {author} {\bibfnamefont {J.~D.}\ \bibnamefont
  {Farmer}}, \bibinfo {author} {\bibfnamefont {A.}~\bibnamefont {Flache}},
  \bibinfo {author} {\bibfnamefont {D.}~\bibnamefont {Garlaschelli}}, \bibinfo
  {author} {\bibfnamefont {A.~G.}\ \bibnamefont {Haldane}}, \bibinfo {author}
  {\bibfnamefont {H.}~\bibnamefont {Heesterbeek}}, \bibinfo {author}
  {\bibfnamefont {C.}~\bibnamefont {Hommes}}, \bibinfo {author} {\bibfnamefont
  {C.}~\bibnamefont {Jaeger}}, \bibinfo {author} {\bibfnamefont
  {R.}~\bibnamefont {May}}, \ and\ \bibinfo {author} {\bibfnamefont
  {M.}~\bibnamefont {Scheffer}},\ }\href@noop {} {\bibfield  {journal}
  {\bibinfo  {journal} {Science}\ }\textbf {\bibinfo {volume} {351}},\ \bibinfo
  {pages} {818} (\bibinfo {year} {2016})}\BibitemShut {NoStop}%
\bibitem [{\citenamefont {Gai}\ and\ \citenamefont
  {Kapadia}(2010)}]{gai2010contagion}%
  \BibitemOpen
  \bibfield  {author} {\bibinfo {author} {\bibfnamefont {P.}~\bibnamefont
  {Gai}}\ and\ \bibinfo {author} {\bibfnamefont {S.}~\bibnamefont {Kapadia}},\
  }\href@noop {} {\bibfield  {journal} {\bibinfo  {journal} {Proceedings of the
  Royal Society A: Mathematical, Physical and Engineering Sciences}\ }\textbf
  {\bibinfo {volume} {466}},\ \bibinfo {pages} {2401} (\bibinfo {year}
  {2010})}\BibitemShut {NoStop}%
\bibitem [{\citenamefont {Beale}\ \emph {et~al.}(2011)\citenamefont {Beale},
  \citenamefont {Rand}, \citenamefont {Battey}, \citenamefont {Croxson},
  \citenamefont {May},\ and\ \citenamefont {Nowak}}]{beale2011individual}%
  \BibitemOpen
  \bibfield  {author} {\bibinfo {author} {\bibfnamefont {N.}~\bibnamefont
  {Beale}}, \bibinfo {author} {\bibfnamefont {D.~G.}\ \bibnamefont {Rand}},
  \bibinfo {author} {\bibfnamefont {H.}~\bibnamefont {Battey}}, \bibinfo
  {author} {\bibfnamefont {K.}~\bibnamefont {Croxson}}, \bibinfo {author}
  {\bibfnamefont {R.~M.}\ \bibnamefont {May}}, \ and\ \bibinfo {author}
  {\bibfnamefont {M.~A.}\ \bibnamefont {Nowak}},\ }\href@noop {} {\bibfield
  {journal} {\bibinfo  {journal} {Proceedings of the National Academy of
  Sciences}\ }\textbf {\bibinfo {volume} {108}},\ \bibinfo {pages} {12647}
  (\bibinfo {year} {2011})}\BibitemShut {NoStop}%
\bibitem [{\citenamefont {Battiston}\ \emph
  {et~al.}(2012{\natexlab{a}})\citenamefont {Battiston}, \citenamefont
  {Puliga}, \citenamefont {Kaushik}, \citenamefont {Tasca},\ and\ \citenamefont
  {Caldarelli}}]{battiston2012debtrank}%
  \BibitemOpen
  \bibfield  {author} {\bibinfo {author} {\bibfnamefont {S.}~\bibnamefont
  {Battiston}}, \bibinfo {author} {\bibfnamefont {M.}~\bibnamefont {Puliga}},
  \bibinfo {author} {\bibfnamefont {R.}~\bibnamefont {Kaushik}}, \bibinfo
  {author} {\bibfnamefont {P.}~\bibnamefont {Tasca}}, \ and\ \bibinfo {author}
  {\bibfnamefont {G.}~\bibnamefont {Caldarelli}},\ }\href@noop {} {\bibfield
  {journal} {\bibinfo  {journal} {Scientific reports}\ }\textbf {\bibinfo
  {volume} {2}} (\bibinfo {year} {2012}{\natexlab{a}})}\BibitemShut {NoStop}%
\bibitem [{\citenamefont {Battiston}\ \emph
  {et~al.}(2012{\natexlab{b}})\citenamefont {Battiston}, \citenamefont {Gatti},
  \citenamefont {Gallegati}, \citenamefont {Greenwald},\ and\ \citenamefont
  {Stiglitz}}]{battiston2012liaisons}%
  \BibitemOpen
  \bibfield  {author} {\bibinfo {author} {\bibfnamefont {S.}~\bibnamefont
  {Battiston}}, \bibinfo {author} {\bibfnamefont {D.~D.}\ \bibnamefont
  {Gatti}}, \bibinfo {author} {\bibfnamefont {M.}~\bibnamefont {Gallegati}},
  \bibinfo {author} {\bibfnamefont {B.}~\bibnamefont {Greenwald}}, \ and\
  \bibinfo {author} {\bibfnamefont {J.~E.}\ \bibnamefont {Stiglitz}},\
  }\href@noop {} {\bibfield  {journal} {\bibinfo  {journal} {Journal of
  economic dynamics and control}\ }\textbf {\bibinfo {volume} {36}},\ \bibinfo
  {pages} {1121} (\bibinfo {year} {2012}{\natexlab{b}})}\BibitemShut {NoStop}%
\bibitem [{\citenamefont {Minoiu}\ and\ \citenamefont
  {Reyes}(2013)}]{minoiu2013network}%
  \BibitemOpen
  \bibfield  {author} {\bibinfo {author} {\bibfnamefont {C.}~\bibnamefont
  {Minoiu}}\ and\ \bibinfo {author} {\bibfnamefont {J.~A.}\ \bibnamefont
  {Reyes}},\ }\href@noop {} {\bibfield  {journal} {\bibinfo  {journal} {Journal
  of Financial Stability}\ }\textbf {\bibinfo {volume} {9}},\ \bibinfo {pages}
  {168} (\bibinfo {year} {2013})}\BibitemShut {NoStop}%
\bibitem [{\citenamefont {Chinazzi}\ \emph {et~al.}(2013)\citenamefont
  {Chinazzi}, \citenamefont {Fagiolo}, \citenamefont {Reyes},\ and\
  \citenamefont {Schiavo}}]{chinazzi2013post}%
  \BibitemOpen
  \bibfield  {author} {\bibinfo {author} {\bibfnamefont {M.}~\bibnamefont
  {Chinazzi}}, \bibinfo {author} {\bibfnamefont {G.}~\bibnamefont {Fagiolo}},
  \bibinfo {author} {\bibfnamefont {J.~A.}\ \bibnamefont {Reyes}}, \ and\
  \bibinfo {author} {\bibfnamefont {S.}~\bibnamefont {Schiavo}},\ }\href@noop
  {} {\bibfield  {journal} {\bibinfo  {journal} {Journal of Economic Dynamics
  and Control}\ }\textbf {\bibinfo {volume} {37}},\ \bibinfo {pages} {1692}
  (\bibinfo {year} {2013})}\BibitemShut {NoStop}%
\bibitem [{\citenamefont {Poledna}\ \emph {et~al.}(2015)\citenamefont
  {Poledna}, \citenamefont {Molina-Borboa}, \citenamefont
  {Mart{\'\i}nez-Jaramillo}, \citenamefont {Van Der~Leij},\ and\ \citenamefont
  {Thurner}}]{poledna2015multi}%
  \BibitemOpen
  \bibfield  {author} {\bibinfo {author} {\bibfnamefont {S.}~\bibnamefont
  {Poledna}}, \bibinfo {author} {\bibfnamefont {J.~L.}\ \bibnamefont
  {Molina-Borboa}}, \bibinfo {author} {\bibfnamefont {S.}~\bibnamefont
  {Mart{\'\i}nez-Jaramillo}}, \bibinfo {author} {\bibfnamefont
  {M.}~\bibnamefont {Van Der~Leij}}, \ and\ \bibinfo {author} {\bibfnamefont
  {S.}~\bibnamefont {Thurner}},\ }\href@noop {} {\bibfield  {journal} {\bibinfo
   {journal} {Journal of Financial Stability}\ }\textbf {\bibinfo {volume}
  {20}},\ \bibinfo {pages} {70} (\bibinfo {year} {2015})}\BibitemShut {NoStop}%
\bibitem [{\citenamefont {Montagna}\ and\ \citenamefont
  {Kok}(2016)}]{montagna2016multi}%
  \BibitemOpen
  \bibfield  {author} {\bibinfo {author} {\bibfnamefont {M.}~\bibnamefont
  {Montagna}}\ and\ \bibinfo {author} {\bibfnamefont {C.}~\bibnamefont {Kok}},\
  }\href@noop {} {\bibfield  {journal} {\bibinfo  {journal} {ECB Working Paper
  No. 1944, Available at SSRN: https://ssrn.com/abstract=2830546}\ } (\bibinfo
  {year} {2016})}\BibitemShut {NoStop}%
\bibitem [{\citenamefont {Brummitt}\ and\ \citenamefont
  {Kobayashi}(2015)}]{brummitt2015cascades}%
  \BibitemOpen
  \bibfield  {author} {\bibinfo {author} {\bibfnamefont {C.~D.}\ \bibnamefont
  {Brummitt}}\ and\ \bibinfo {author} {\bibfnamefont {T.}~\bibnamefont
  {Kobayashi}},\ }\href@noop {} {\bibfield  {journal} {\bibinfo  {journal}
  {Physical Review E}\ }\textbf {\bibinfo {volume} {91}},\ \bibinfo {pages}
  {062813} (\bibinfo {year} {2015})}\BibitemShut {NoStop}%
\bibitem [{\citenamefont {Poledna}\ \emph {et~al.}(2018)\citenamefont
  {Poledna}, \citenamefont {Hinteregger},\ and\ \citenamefont
  {Thurner}}]{poledna2018identifying}%
  \BibitemOpen
  \bibfield  {author} {\bibinfo {author} {\bibfnamefont {S.}~\bibnamefont
  {Poledna}}, \bibinfo {author} {\bibfnamefont {A.}~\bibnamefont
  {Hinteregger}}, \ and\ \bibinfo {author} {\bibfnamefont {S.}~\bibnamefont
  {Thurner}},\ }\href@noop {} {\bibfield  {journal} {\bibinfo  {journal} {arXiv
  preprint arXiv:1801.10487}\ } (\bibinfo {year} {2018})}\BibitemShut {NoStop}%
\bibitem [{\citenamefont {Zlati{\'c}}\ \emph {et~al.}(2015)\citenamefont
  {Zlati{\'c}}, \citenamefont {Gabbi},\ and\ \citenamefont
  {Abraham}}]{zlatic2015reduction}%
  \BibitemOpen
  \bibfield  {author} {\bibinfo {author} {\bibfnamefont {V.}~\bibnamefont
  {Zlati{\'c}}}, \bibinfo {author} {\bibfnamefont {G.}~\bibnamefont {Gabbi}}, \
  and\ \bibinfo {author} {\bibfnamefont {H.}~\bibnamefont {Abraham}},\
  }\href@noop {} {\bibfield  {journal} {\bibinfo  {journal} {PloS one}\
  }\textbf {\bibinfo {volume} {10}},\ \bibinfo {pages} {e0114928} (\bibinfo
  {year} {2015})}\BibitemShut {NoStop}%
\bibitem [{\citenamefont {Poledna}\ and\ \citenamefont
  {Thurner}(2016)}]{poledna2016elimination}%
  \BibitemOpen
  \bibfield  {author} {\bibinfo {author} {\bibfnamefont {S.}~\bibnamefont
  {Poledna}}\ and\ \bibinfo {author} {\bibfnamefont {S.}~\bibnamefont
  {Thurner}},\ }\href@noop {} {\bibfield  {journal} {\bibinfo  {journal}
  {Quantitative Finance}\ }\textbf {\bibinfo {volume} {16}},\ \bibinfo {pages}
  {1599} (\bibinfo {year} {2016})}\BibitemShut {NoStop}%
\bibitem [{\citenamefont {Krause}\ \emph {et~al.}(2021)\citenamefont {Krause},
  \citenamefont {{\v{S}}tefan{\v{c}}i{\'c}}, \citenamefont {Caldarelli},\ and\
  \citenamefont {Zlati{\'c}}}]{krause2021controlling}%
  \BibitemOpen
  \bibfield  {author} {\bibinfo {author} {\bibfnamefont {S.~M.}\ \bibnamefont
  {Krause}}, \bibinfo {author} {\bibfnamefont {H.}~\bibnamefont
  {{\v{S}}tefan{\v{c}}i{\'c}}}, \bibinfo {author} {\bibfnamefont
  {G.}~\bibnamefont {Caldarelli}}, \ and\ \bibinfo {author} {\bibfnamefont
  {V.}~\bibnamefont {Zlati{\'c}}},\ }\href@noop {} {\bibfield  {journal}
  {\bibinfo  {journal} {Physical Review E}\ }\textbf {\bibinfo {volume}
  {103}},\ \bibinfo {pages} {042304} (\bibinfo {year} {2021})}\BibitemShut
  {NoStop}%
\bibitem [{\citenamefont {Roukny}\ \emph {et~al.}(2013)\citenamefont {Roukny},
  \citenamefont {Bersini}, \citenamefont {Pirotte}, \citenamefont
  {Caldarelli},\ and\ \citenamefont {Battiston}}]{roukny2013default}%
  \BibitemOpen
  \bibfield  {author} {\bibinfo {author} {\bibfnamefont {T.}~\bibnamefont
  {Roukny}}, \bibinfo {author} {\bibfnamefont {H.}~\bibnamefont {Bersini}},
  \bibinfo {author} {\bibfnamefont {H.}~\bibnamefont {Pirotte}}, \bibinfo
  {author} {\bibfnamefont {G.}~\bibnamefont {Caldarelli}}, \ and\ \bibinfo
  {author} {\bibfnamefont {S.}~\bibnamefont {Battiston}},\ }\href@noop {}
  {\bibfield  {journal} {\bibinfo  {journal} {Scientific reports}\ }\textbf
  {\bibinfo {volume} {3}},\ \bibinfo {pages} {2759} (\bibinfo {year}
  {2013})}\BibitemShut {NoStop}%
\bibitem [{\citenamefont {Bardoscia}\ \emph {et~al.}(2015)\citenamefont
  {Bardoscia}, \citenamefont {Battiston}, \citenamefont {Caccioli},\ and\
  \citenamefont {Caldarelli}}]{bardoscia2015}%
  \BibitemOpen
  \bibfield  {author} {\bibinfo {author} {\bibfnamefont {M.}~\bibnamefont
  {Bardoscia}}, \bibinfo {author} {\bibfnamefont {S.}~\bibnamefont
  {Battiston}}, \bibinfo {author} {\bibfnamefont {F.}~\bibnamefont {Caccioli}},
  \ and\ \bibinfo {author} {\bibfnamefont {G.}~\bibnamefont {Caldarelli}},\
  }\href@noop {} {\bibfield  {journal} {\bibinfo  {journal} {PloS one}\
  }\textbf {\bibinfo {volume} {10}},\ \bibinfo {pages} {e0130406} (\bibinfo
  {year} {2015})}\BibitemShut {NoStop}%
\bibitem [{\citenamefont {Gibson}()}]{dow}%
  \BibitemOpen
  \bibfield  {author} {\bibinfo {author} {\bibfnamefont {K.}~\bibnamefont
  {Gibson}},\ }\href
  {https://www.marketwatch.com/story/us-stocks-slide-dow-plunges-777-points-as-bailout-bill-fails-2008929164700}
  {\enquote {\bibinfo {title} {U.s. stocks hammered after house rejects
  rescue},}\ }\BibitemShut {NoStop}%
\bibitem [{\citenamefont {Kwapisz}(1996)}]{kwapisz1996spectral}%
  \BibitemOpen
  \bibfield  {author} {\bibinfo {author} {\bibfnamefont {J.}~\bibnamefont
  {Kwapisz}},\ }\href@noop {} {\bibfield  {journal} {\bibinfo  {journal}
  {Journal of graph theory}\ }\textbf {\bibinfo {volume} {23}},\ \bibinfo
  {pages} {405} (\bibinfo {year} {1996})}\BibitemShut {NoStop}%
\bibitem [{\citenamefont {Chalkis}\ and\ \citenamefont
  {Fisikopoulos}(2020)}]{chalkis2020volesti}%
  \BibitemOpen
  \bibfield  {author} {\bibinfo {author} {\bibfnamefont {A.}~\bibnamefont
  {Chalkis}}\ and\ \bibinfo {author} {\bibfnamefont {V.}~\bibnamefont
  {Fisikopoulos}},\ }\href@noop {} {\bibfield  {journal} {\bibinfo  {journal}
  {arXiv preprint arXiv:2007.01578}\ } (\bibinfo {year} {2020})}\BibitemShut
  {NoStop}%
\end{thebibliography}%

\end{document}